\documentclass[12pt]{article}
\usepackage{arxiv}
\usepackage{graphicx}
\usepackage{amsmath}
\usepackage{apacite}
\usepackage{booktabs}
\usepackage{mathtools}
\usepackage{algorithm}
\usepackage[noend]{algpseudocode}

\newcommand{\beq}{\begin{equation}}
\newcommand{\eeq}{\end{equation}}
\newcommand{\bea}{\begin{eqnarray}}
\newcommand{\eea}{\end{eqnarray}}
\newcommand{\ba}{\begin{array}}
\newcommand{\ea}{\end{array}}
\newcommand{\bi}{\begin{itemize}}
\newcommand{\ei}{\end{itemize}}
\newcommand{\ben}{\begin{enumerate}}
\newcommand{\een}{\end{enumerate}}

\mathtoolsset{showonlyrefs=true}
\usepackage{amsmath,amsfonts,amssymb}
\usepackage{apacite}
\title{Bayesian Benefit–Risk Assessment with Dependent Outcomes via Latent Factor Models}

\author{
 Konstantinos Vamvourellis, Konstantinos Kalogeropoulos, Lawrence Phillips\\
 Department of Statistics, LSE\\
  \texttt{k.vamvourellis@lse.ac.uk}; \texttt{k.kalogeropoulos@lse.ac.uk;}\\
}
\begin{document}
\maketitle
\begin{abstract}
Approving and assessing new drugs is complex because multiple criteria must be considered simultaneously. A common approach is benefit–risk analysis, often conducted within a Bayesian framework to account for uncertainty and combine data with expert judgement, typically through multi-criteria decision analysis (MCDA) scores. This requires models that accommodate mixed and potentially correlated outcomes; latent factor models provide a natural framework.
We develop a coherent Bayesian framework for benefit–risk analysis that addresses these challenges and supports sequential decision-making. We extend structured factor models to mixed outcomes and introduce a principled approach for selecting among competing specifications that combines model fit with out-of-sample predictive performance.
We then develop a sequential estimation framework that updates MCDA scores as new data become available, allowing treatment comparisons to evolve over time. This supports early stopping when conclusions are clear and permits dynamic treatment allocation aligned with study objectives. To make this feasible, we develop tailored sequential Monte Carlo methods adapted to the model structure.
The methodology is illustrated using data on patients with type II diabetes treated with Metformin, Rosiglitazone, and their combination.
\end{abstract}


\section{Introduction}

Regulatory authorities responsible for authorising new drugs often face a difficult challenge when choosing an appropriate course of action in the presence of multiple competing objectives. Rising demand for individualised treatments, the increasing number of stakeholders, and the growing complexity of the questions posed in modern clinical settings all contribute to this difficulty. Multiple high-profile drugs have been withdrawn over the past 20 years \cite{guo2010review}, some permanently, while others have been remarketed under revised labelling and guidance \cite{WWZC20}. Given both the complexity and the consequences of such decisions, there is a clear need for structured quantitative approaches to benefit--risk assessment of medicines. Such approaches, including the framework developed in this paper, typically rely on modelling clinical trial or observational data to estimate relevant quantities while balancing competing objectives.

One prominent example is Rosiglitazone, a treatment for type II diabetes marketed under the commercial name Avandia. It gained market authorisation in the United States in 1999 and in the European Union in 2000. New data subsequently emerged about possible cardiovascular risks associated with Rosiglitazone, confirmed by a meta-analysis in 2007 \cite{nissen2007effect}, which resulted in a European suspension of the marketing authorisation in 2010. This suspension included its use as a fixed dose combination with Metformin or Glimepiride for type II diabetes, which had been approved in 2003 for Metformin and 2006 for Glimepiride. The drug remained available in the United States, but only under a restricted-access programme put in place in 2010. In 2011 the US regulators revised the recommendation and relaxed these restrictions, and in 2013 the drug regained approval following a study that found Rosiglitazone to be as safe as other diabetes drugs. Today the drug is withdrawn from most countries in the world, but remains available in the United States. For a complete review of the regulatory history of Rosiglitazone see \shortciteA{WWZC20}.

Benefit--risk analysis is an umbrella term that encompasses any structured approach for weighing the benefits (typically treatment efficacy) against the risks (typically adverse or unwanted side effects) of drug treatments. The goal is to assist stakeholders in deciding whether a drug is effective and whether its potential side effects are acceptable. Since such decisions depend on individual preferences and risk tolerances, proposed frameworks often incorporate formal or informal utility functions. \shortciteA{mt2014balancing} review the benefit--risk literature and categorise approaches into quantitative frameworks, metrics for benefit--risk assessment, estimation and utility survey techniques, and qualitative frameworks. Of these, the first two correspond to formal quantitative approaches grounded in statistical methodology, whereas the latter two are more informal. Common single metrics include the benefit--risk ratio \shortcite{carlisle2016benefit} and net clinical benefit \shortcite{shakespeare2001improving}; see also \citeA{holden2003benefit}. While such metrics provide useful summaries, they are limited in their ability to capture the full complexity of decision-making, as they typically focus on specific aspects of the problem. In practice, they often require benefits and risks to be expressed on a common scale, which may not be straightforward. From a theoretical perspective, complex decisions cannot always be reduced to a single number, particularly when uncertainty is present. Ratio-based metrics, in particular, have been criticised for instability and for obscuring uncertainty in the final decision \shortcite{lynd2004advances, shaffer2006joint, sutton2005bayesian}.

Multi-criteria decision analysis (MCDA) provides a flexible and comprehensive quantitative framework for benefit--risk assessment and is widely used in practice \shortcite{keeney1993decisions, mussen2009benefit, phillips2014}. It allows outcomes to be weighted according to their utilities and supports the construction of integrated summary scores. MCDA has been developed over several decades \cite{jong1976keeney} and has been applied in the evaluation of drug treatments \cite{glasziou1995evidence} as well as other interventions \cite{ponce2000use}. It has also been recommended for use in regulatory settings \shortcite{mussen2007quantitative, garrison2007assessing, muhlbacher2016patient}. In addition, MCDA lends itself naturally to sensitivity analysis with respect to key inputs, leading to methodologies such as stochastic multi-criteria acceptability analysis \shortcite{tervonen2008survey, tervonen2011stochastic, lahdelma1998smaa}, for which software implementations are available \cite{tervonen2014jsmaa}.

Recent work has explored the use of Bayesian modelling within MCDA frameworks. \citeA{waddingham2016bayesian} consider a Bayesian probabilistic model that propagates sampling uncertainty into the final MCDA score using a combination of aggregated and patient-level data. Their approach, however, assumes independence across outcomes and does not address model assessment. \citeA{li2019bayesian} propose a latent trait model that accounts for correlation between outcomes and accommodates both continuous and binary data, introducing latent factors representing benefits and risks.

These contributions demonstrate the value of Bayesian modelling for incorporating uncertainty and modelling dependence in benefit--risk analysis. However, several challenges remain. Existing approaches either rely on independence assumptions or consider specific latent structures without providing a systematic framework for assessing model adequacy. Because final decisions are based on model outputs, it is important to evaluate whether a proposed model is supported by the data. Moreover, predictive performance plays a key role when choosing between competing specifications, particularly in settings where decisions may need to be updated as new data become available.

In this paper, we develop a coherent framework for Bayesian benefit--risk analysis that addresses these challenges and supports sequential decision-making. We extend existing factor models to accommodate mixed-type and potentially dependent outcomes using a Bayesian structural equation modelling approach \shortcite{muthen2012, vamvourellis2021generalised}, and consider both treatment-specific and pooled specifications. This gives rise to a set of competing models, highlighting the need for principled model assessment. We therefore introduce a framework for model selection that balances goodness-of-fit with predictive performance. Building on this, we consider a sequential clinical study design in which MCDA scores are updated as new data become available. This enables monitoring of treatment differences over time, supports early stopping when conclusions become clear, and allows dynamic allocation of patients across treatment groups based on the evolving evidence. To support this sequential framework, we develop tailored Sequential Monte Carlo methods that efficiently carry out the required posterior updates. To our knowledge, this is among the first applications of such methods in the context of benefit--risk analysis.

The paper is structured as follows. In Section 2, we present the modelling framework for benefit--risk analysis, including the data-generating process and the MCDA formulation. We review alternative latent variable models, including exploratory factor analysis and structural equation models, to capture different features of the data. Section 3 introduces the framework for model assessment and selection. Section 4 develops the sequential inference approach. In Section 5, we illustrate the proposed methodology using a clinical case study involving treatments for type II diabetes, including Rosiglitazone, Metformin, and Glimepiride. We conclude with a discussion of limitations and directions for future work.

\section{Factor Models for Multi-criteria Decision Analysis}\label{section:methods}

\subsection{Multi-criteria Decision Analysis Score and Data}

We consider an MCDA framework for $R$ treatments, based on data collected on $P$ criteria for a set of $N$ subjects. Let $y_{ijr}$ denote the observation for subject $i=1,\dots,N$, criterion $j=1,\dots,P$, and treatment $r=1,\dots,R$. In a clinical trial setting, the criteria typically consist of treatment benefits and adverse events. Benefits are often measured through continuous efficacy outcomes, whereas adverse events are commonly recorded as counts or binary indicators of occurrence. The MCDA score is constructed from summaries of the outcomes across subjects. In particular, it requires the expected value of the outcome for criterion $j$ under treatment $r$, denoted by $\mu_{jr} := E(y_{ijr})$, where the expectation is taken with respect to the data-generating distribution implied by a statistical model. In this paper, we focus on models where this quantity represents the mean outcome for criterion $j$ under treatment $r$ at the population level, but extensions to individual-specific quantities can be accommodated within the same framework.

For each criterion $j$, a utility function $U_j(\cdot): \mathbb{R} \to [0,1]$ maps the expected outcome to a common benefit--risk scale, where $0$ represents the least favourable outcome and $1$ the most favourable. These functions are typically specified a priori and are taken to be common across subjects and treatments. In this paper, we consider simple linear mappings, although more general transformations can also be accommodated within our framework. Finally, each criterion is assigned a weight $w_j$ reflecting its relative importance, with $\sum_{j=1}^P w_j = 1$. These weights may be elicited from expert clinicians or individual patients. For simplicity, we assume common weights across subjects, although subject-specific weights ${w_{ij}}$ could also be incorporated.

The population-based MCDA score for treatment $r$ is then defined as
\begin{equation}
\label{eq:MCDA}
M_r := \sum_{j=1}^P w_j U_{j}(\mu_{jr}).
\end{equation}
Evaluation of $M_r$ therefore requires specifying a statistical model. Let $\Theta$ denote the parameters of such a model, and let $Y = \{ y_{ijr} : i=1,\dots,N,\; j=1,\dots,P,\; r=1,\dots,R \}$ denote the observed data. Under this model, the quantities $\mu_{jr}$ are functions of $\Theta$, and hence the MCDA score can be written as $M_r(\Theta)$.

Under a frequentist approach, one may estimate $\hat{\Theta}$ from $Y$ and compute $M_r(\hat{\Theta})$, possibly accompanied by inference on differences between treatments. Under a Bayesian framework, given a prior $\pi(\Theta)$, inference proceeds via the posterior distribution $\pi\big(M_r(\Theta)\mid Y\big)$, allowing quantities such as the posterior probability that one treatment has a higher MCDA score than another to be evaluated.

In either framework, several sources of variability arise: uncertainty due to the observed data $Y$, uncertainty in the model parameters, and uncertainty associated with the choice of model. In the remainder of this section and the next, we develop a Bayesian framework for benefit--risk analysis that explicitly accounts for these components and propagates them to the resulting MCDA scores.

\subsection{Factor Analysis for Mixed-Type Data}
\label{sec:models}

Since the MCDA score in \eqref{eq:MCDA} depends on unknown population means $\mu_{jr}$, these quantities must be estimated from an appropriate statistical model. The data consist of observations $\{y_{ijr}\}$ for subjects $i=1,\dots,N$, 
criteria $j=1,\dots,P$, and treatments $r=1,\dots,R$. 
In this section, we focus on modelling a single treatment group and suppress 
the treatment index $r$, writing $y_{ij}$ in place of $y_{ijr}$ for notational simplicity. The models considered in this section build on well-established factor analytic and structural equation modelling frameworks, including exploratory and confirmatory factor models, as well as their Bayesian extensions \shortcite{muthen2012, vamvourellis2021generalised}. Our focus is on adapting and integrating these models within a unified framework for benefit--risk analysis with mixed-type outcomes, and on assessing their suitability for decision-making purposes.

\subsubsection{Latent variable framework}

To model mixed-type and potentially dependent outcomes, we adopt a latent variable approach based on factor analysis. For each subject $i$, we introduce a vector of $k$ latent factors $\mathbf{z}_i = (z_{i1}, \dots, z_{ik})$. These factors capture dependence across the observed criteria. We further introduce latent continuous variables $y^*_{ij}$ for each observed outcome $y_{ij}$. 
The relationship between latent variables and factors is given by
\begin{equation}
\label{augmented}
y^*_{ij} = \alpha_j + \lambda_j^\top \mathbf{z}_i + \epsilon_{ij}, 
\quad i=1,\dots,N,\; j=1,\dots,P,
\end{equation}
where $\alpha_j$ is an intercept, $\lambda_j$ is a $k$-dimensional vector of loadings (the $j$-th row of $\Lambda$), and $\epsilon_{ij}$ is an error term.

The observed data $y_{ij}$ are linked to the latent variables $y^*_{ij}$ as follows. 
If $y_{ij}$ is continuous, we assume $y_{ij} = y^*_{ij}$. 
If $y_{ij}$ is binary, we assume $y_{ij} = \mathcal{I}(y^*_{ij} > 0)$, where $\mathcal{I}(\cdot)$ denotes the indicator function.  
If $y_{ij}$ is ordinal with $m_j$ categories, we assume
\[
y_{ij} = a \quad \text{if} \quad \tau^{(j)}_{a-1} < y^*_{ij} \le \tau^{(j)}_{a}, 
\quad a = 1, \dots, m_j,
\]
where $\{\tau^{(j)}_a\}$ are threshold parameters, with $\tau^{(j)}_0 = -\infty$ and $\tau^{(j)}_{m_j} = +\infty$.

We assume $\mathbf{z}_i \sim N_k(0, \Phi)$ and $\epsilon_{ij} \sim N(0, \psi_j^2)$, independently across $i$ and $j$. 
This formulation provides a unified framework for modelling continuous and categorical data.

In the case where all outcomes are continuous, the latent variables can be marginalised to obtain
\begin{equation}
\label{marginal}
\mathbf{y}_i \sim N(\alpha, \Lambda \Phi \Lambda^\top + \Psi), \quad i=1,\dots,N,
\end{equation}
where $\Psi = \mathrm{diag}(\psi_1^2,\dots,\psi_P^2)$.

When the loading structure $\Lambda$ is constrained based on prior knowledge, the model corresponds to confirmatory factor analysis (CFA). When no such constraints are imposed, the model corresponds to exploratory factor analysis (EFA), typically requiring additional identifiability constraints such as $\Phi = I_k$. A limiting case is the saturated model
\begin{equation}
\label{eq:saturated}
\mathbf{y}_i \sim N(\alpha, \Sigma),
\end{equation}
where $\Sigma$ is an unrestricted covariance matrix (subject to standard identifiability constraints for categorical variables).
\subsubsection{Bayesian structural equation modelling}
\label{sec:AZ}

The Bayesian formulation of factor models, introduced by \shortciteA{muthen2012} and extended by \shortciteA{vamvourellis2021generalised}, provides an alternative specification by replacing exact zero restrictions on elements of $\Lambda$ with approximate ones. This is achieved by assigning informative priors that concentrate mass near zero; we refer to such models as \emph{approximate zero} models. The latent variable framework in \eqref{augmented} is extended by introducing variable-specific random effects $u_{ij}$, leading to
\begin{equation}
\label{newmodel}
y^*_{ij} = \alpha_j + \lambda_j^\top \mathbf{z}_i + u_{ij} + \epsilon_{ij}, 
\quad i=1,\dots,N,\; j=1,\dots,P,
\end{equation}
where $\mathbf{u}_i = (u_{i1}, \dots, u_{iP})^\top$ is a vector of random effects with covariance matrix $\Omega$, and $\epsilon_{ij}$ remains an error term with variance $\psi_j^2$. The additional random effects $u_{ij}$ capture residual associations between outcomes that are not explained by the latent factors $\mathbf{z}_i$, typically of smaller magnitude. We assume $\mathbf{u}_i \sim N(0,\Omega)$, $\mathbf{z}_i \sim N(0,\Phi)$, and $\epsilon_{ij} \sim N(0,\psi_j^2)$, independently across $i$ and $j$. In the case of continuous data, the latent variables can be marginalised to obtain
\begin{equation}
\label{newmarginal}
\mathbf{y}_i \sim N\big(\alpha, \Lambda \Phi \Lambda^\top + \Omega + \Psi \big), 
\quad i=1,\dots,N,
\end{equation}
where $\Psi = \mathrm{diag}(\psi_1^2,\dots,\psi_P^2)$.

Compared to the model in \eqref{marginal}, this formulation relaxes the diagonal structure of $\Psi$ by allowing for additional dependence through $\Omega$, while controlling its magnitude via informative priors. As such, the approximate zero model can be viewed as an intermediate specification between confirmatory and exploratory factor analysis.

\subsubsection{Multiple group models}

The models presented so far refer to a single treatment group. When data from multiple groups are available, one can either fit separate models for each group or consider partially pooled specifications to improve parsimony.

Suppose subjects are allocated to one of $R$ treatment groups, and let $\{\mathbf{y}^{(r)}_i\}_{i=1}^{n_r}$ denote the data for group $r$, where $n_r$ is the number of subjects in that group with $\sum_{r=1}^R n_r=N$. Rather than modelling each group independently, it may be beneficial to share certain parameters across groups, such as the covariance components $(\Phi, \Omega, \Psi)$, while allowing others, such as the intercepts $\alpha_r$ and loadings $\Lambda_r$, to vary by group.

In this case, the marginal model for continuous data becomes
\begin{equation}
\label{poolmarginal}
\mathbf{y}^{(r)}_i \sim N\big(\alpha_r, \Lambda_r \Phi \Lambda_r^\top + \Omega + \Psi \big), 
\quad i=1,\dots,n_r,\; r=1,\dots,R.
\end{equation}

\subsection{Bayesian specification}
\label{sec:priors}

The Bayesian specification is completed by assigning prior distributions to the unknown parameters. Inference is carried out using MCMC methods, with implementation details provided in Section S1 of the Supplementary Material. The priors are chosen to provide weakly informative regularisation while preserving flexibility in the latent structure. The resulting posterior distribution combines the likelihood with these prior specifications. For the loading matrix $\Lambda$, we follow standard recommendations in the literature. The principal (non-zero) loadings are assigned Gaussian priors, $\Lambda_{ij} \sim N(0,\sigma_j^2)$ for continuous outcomes \shortcite{CFHP14, FL18}, and $\Lambda_{ij} \sim N(0,4)$ for binary outcomes \shortcite{VNM14}. Under the approximate zero framework, cross-loadings are assigned more concentrated priors, $\Lambda_{ij} \sim N(0,0.1^2)$, as in \citeA{muthen2012}. For continuous variables, the idiosyncratic variances are assigned inverse-gamma priors
\[
\psi_j^2 \sim \text{InvGamma}\!\left(c_0, \frac{c_0-1}{(S_y^{-1})_{jj}}\right),
\]
where $S_y$ is the sample covariance matrix of the continuous observations and $c_0$ is chosen to avoid degeneracies such as Heywood cases. The latent factor covariance matrix $\Phi$ is parameterised as a correlation matrix, and an LKJ prior \shortcite{LKJ09} is assigned to it. For identification, restrictions are imposed on the loading matrix; rather than fixing diagonal elements, we follow a parameter expansion approach similar to \citeA{GD09}, assigning Gaussian priors and applying post-processing to enforce identifiability. For the residual covariance matrix $\Omega$, we use an inverse-Wishart prior with identity scale matrix and $P+6$ degrees of freedom, reflecting a prior belief in small residual correlations \shortcite{muthen2012}. Finally, intercept parameters are assigned diffuse Gaussian priors, $\alpha_j \sim N(0,10^2)$. 

\section{Model Assessment and Selection for Benefit--Risk Inference}
\label{section:modelfit}

\subsection{Candidate model specifications}

The modelling framework of Section \ref{sec:models} gives rise to a range of plausible specifications, reflecting different assumptions on the latent structure and dependence across outcomes. In particular, competing models may differ in the number of latent factors, the structure of the loading matrix (e.g. restricted versus unrestricted cross-loadings), and the degree of sparsity imposed on secondary loadings through prior distributions. Simpler specifications, including models that assume conditional independence across outcomes, may also be considered as benchmarks. While more flexible models can capture complex dependence patterns across efficacy and safety endpoints, they may also lead to overfitting and reduced predictive performance. Conversely, overly restrictive models may fail to capture clinically meaningful associations between outcomes, potentially affecting the estimation of the population means $\mu_{jr}$ and, in turn, the resulting MCDA scores in \eqref{eq:MCDA}. This motivates the need for a principled model assessment and selection framework that balances goodness-of-fit and predictive performance. We adopt the general approach of \citeA{vamvourellis2021generalised}, developed in the context of Bayesian structural equation models, and extend it to accommodate the mixed-type outcomes typically encountered in benefit--risk analysis.

\subsection{Model assessment via posterior predictive checks}

Posterior predictive p-values (PPP values) are widely used in Bayesian model assessment \cite{M94}. In the present setting, we adopt this approach and tailor it to the mixed-type outcome framework considered here.

In the context of benefit--risk analysis, model assessment serves not only to evaluate statistical fit, but also to ensure that dependence across outcomes is adequately captured so that uncertainty is correctly propagated to downstream benefit--risk summaries. PPP values provide an absolute measure of in-sample fit by comparing the observed data $Y$ to replicated datasets $\tilde{Y}$ generated from the fitted model. A key component is the discrepancy function $D(Y,\Theta)$, which measures the extent to which the model, under parameter value $\Theta$, deviates from the observed data. Given posterior samples $\{\Theta_m\}_{m=1}^M$, we approximate the PPP value as follows:
\begin{enumerate}
\item For each $m=1,\dots,M$:
\begin{enumerate}
\item Compute $D(Y, \Theta_m)$.
\item Draw a replicated dataset $\tilde{Y}$ from the model using $\Theta_m$, based on \eqref{augmented} or \eqref{newmarginal}.
\item Compute $D(\tilde{Y}, \Theta_m)$ and set 
\[
d_m = \mathcal{I}\big[ D(Y, \Theta_m) < D(\tilde{Y}, \Theta_m) \big].
\]
\end{enumerate}
\item Return $\mathrm{PPP} = \frac{1}{M}\sum_{m=1}^M d_m$.
\end{enumerate}
PPP values are not are not uniformly calibrated frequentist p-values and are therefore interpreted as descriptive measures of fit. In practice, extreme values may indicate that the observed data are atypical under the fitted model in aspects targeted by the chosen discrepancy function. A key challenge in applying PPP-based diagnostics in benefit--risk settings is that outcomes are typically of mixed-type (e.g. continuous continuous measures of treatment benefit and binary adverse events). Standard discrepancy functions do not extend naturally to settings with mixed outcome types. We therefore evaluate PPP values separately by data type, following \citeA{moustaki1996latent}, which allows targeted assessment of different components of the model.

For continuous outcomes, we use the likelihood ratio statistic comparing the factor model to a saturated model \cite<see e.g.>{scheines99test}:
\begin{equation}
\label{gbsem_LRT}
D(Y,\Theta)=(N-1)\left\{\log \left|\Sigma(\Theta)\right| + \mathrm{tr}\left[S(Y)\Sigma^{-1}(\Theta)\right] - \log\left|S(Y)\right| - P \right\},
\end{equation}
where $S(Y)$ and $\Sigma(\Theta)$ denote the sample and model-implied covariance matrices respectively, $N$ is the sample size and $P$ is the number of items. For binary outcomes, it is convenient to work with response patterns. Let $\mathbf{y}_h$, $h=1,\ldots,H$, denote the $H=2^P$ possible response patterns, with observed frequencies $O_h$. The model-implied probability of a response pattern is
\begin{equation}
\label{eq:expected}
\pi_h(\Theta) = \int \prod_{j=1}^P \text{Bernoulli}\left( [\mathbf{y}_h]_j \mid \sigma(\eta_j) \right) f(\mathbf{z}) f(\mathbf{u}) \, d\mathbf{z} \, d\mathbf{u},
\end{equation}
where $\sigma(\cdot)$ denotes the inverse-link function (e.g. logistic or probit). The integral can be approximated using Monte Carlo. A commonly used discrepancy function is the $G^2$ statistic \cite{sinharay05g2}:
\begin{equation}
\label{eq:g2}
D(Y,\Theta) = \sum_{h=1}^H O_h \log\left(\frac{O_h}{N \pi_h(\Theta)}\right),
\end{equation}
which captures discrepancies in higher-order dependence structures across binary outcomes.

\subsection{Predictive performance and model comparison}

While posterior predictive checks assess whether a model can reproduce key features of the observed data, they do not provide a basis for selecting among competing models. For this purpose, we use proper scoring rules, focusing on the log score \cite{dawid2015bayesian,GR07}, following \citeA{vamvourellis2021generalised}.

Let $Y^{tr}$ and $Y^{te}$ denote training and test samples, respectively. The predictive distribution under a model is given by
\begin{equation}
\label{gbsem_eq:predictive}
f(Y^{te} \mid Y^{tr}) = \int f(Y^{te} \mid \Theta)\pi(\Theta \mid Y^{tr}) d\Theta.
\end{equation}
The log score is defined as
\begin{equation}
\label{gbsem_eq:lsrule}
LS(Y^{te}) = - \log f(Y^{te} \mid Y^{tr}),
\end{equation}
with lower values indicating better predictive performance. The log score is strictly proper, providing a principled basis for model comparison.

In our setting, the key challenge is again the presence of mixed-type outcomes. Under the models of Section \ref{sec:models}, the predictive density factorises into continuous and binary components conditional on $\Theta$. We exploit this structure to obtain a tractable approximation of the log score using the mixtures-of-parameters (MP) approach; see, for example, \shortciteA{KLTG21}. Given posterior samples $\{\Theta_m\}_{m=1}^M$ from $\pi(\Theta \mid Y^{tr})$, the log score for an individual observation $Y^{te}_i$ is approximated as
\begin{equation}
LS(Y^{te}_i) \approx -\log \left\{ \frac{1}{M}\sum_{m=1}^M f(Y^{te}_i \mid \Theta_m) \right\}.
\end{equation}

For mixed-type data, writing $Y_i = (C_i, B_i)$ for continuous and binary components, we obtain
\begin{equation}
LS(Y^{te}_i) = - \log \int f(C^{te}_i \mid \Theta) f(B^{te}_i \mid \Theta)\pi(\Theta \mid Y^{tr}) d\Theta,
\end{equation}
with MP approximation
\begin{equation}
LS(Y^{te}_i) \approx - \log \left\{ \frac{1}{M} \sum_{m=1}^M N(C^{te}_i \mid M_c(\Theta_m), V_c(\Theta_m)) \times \pi_h(\Theta_m) \right\}.
\end{equation}

These criteria are used to compare alternative latent factor specifications, including different numbers of factors and loading structures, and to select a model that achieves a balance between flexibility and predictive accuracy.

\subsection{Implications for benefit--risk evaluation}

This section adapts standard model assessment tools to the mixed-type outcome setting of benefit--risk analysis, where both covariance-based and response-pattern-based features must be accommodated within a unified framework. The choice of latent factor specification has direct implications for benefit--risk evaluation, as it determines how dependence across outcomes is modelled and how uncertainty is propagated to summary measures. In particular, different model specifications may lead to different estimates of the population means $\mu_{jr}$, and hence to different MCDA scores $M_r(\Theta)$. In Section \ref{sec:casestudy}, we illustrate how the proposed assessment framework guides model selection in practice and supports benefit--risk evaluation in a clinically relevant setting.

\section{Sequential Benefit--Risk Analysis}
\label{section:smc}

\subsection{Motivation}

As discussed in Section \ref{section:methods}, the MCDA score depends both on the assumed model and on the data used to estimate its parameters. As additional data become available, posterior distributions are updated, leading to changes in the estimated population means $\mu_{jr}$ and, consequently, in the MCDA scores $M_r(\Theta)$. This suggests a sequential perspective on benefit--risk analysis, where inference is updated as data accrue. Such an approach allows monitoring of the evolution of MCDA scores over time, as well as differences between treatments, providing insight into the stability of conclusions and the accumulation of evidence.

To formalise this setting, we impose an ordering on the data, writing $Y_{1:t} = \{y_1,\dots,y_t\}$ for observations up to time $t$, where $t=1,\dots,N$. The ordering may correspond to the chronological order of data collection when available. While the final posterior $\pi(\Theta \mid Y_{1:N})$ is invariant to this ordering, the sequence of intermediate posteriors $\pi(\Theta \mid Y_{1:t})$ is informative about the learning process. Under the Bayesian framework, sequential updating yields a sequence of posteriors $\pi(\Theta \mid Y_{1:t})$ and corresponding distributions for the MCDA scores, $\pi\big(M_r(\Theta) \mid Y_{1:t}\big)$, which form the basis for dynamic benefit--risk assessment.

\subsection{Sequential Monte Carlo for Benefit--Risk Analysis}

Sequential Monte Carlo (SMC) methods provide a natural framework for approximating the sequence of posteriors $\pi(\Theta \mid Y_{1:t})$. The key idea is to represent each posterior by a weighted set of particles and update these sequentially as new observations become available. We adopt a data-tempering approach, constructing a sequence of distributions corresponding to partial posteriors $\pi(\Theta \mid Y_{1:t})$, $t=1,\dots,N$. This is particularly useful to benefit--risk analysis, as it enables direct tracking of how MCDA scores evolve with incoming data. The choice of SMC scheme depends on the structure of the model. When latent variables can be marginalised, inference can be based on the standard IBIS algorithm, which sequentially reweights and rejuvenates parameter particles as new observations arrive, summarised in the Supplementary Section S2.

In more general settings, including the mixed-type models of Section \ref{sec:models}, inference relies on the augmented posterior over parameters and latent variables, denoted by $\Theta$, requiring extensions of IBIS or more general SMC schemes. Regarding the MCDA scores, two cases arise. For population-level scores, the quantities $\mu_{jr}$ are typically functions of the model static parameters $\theta$, and inference can be based on $\pi(\theta \mid Y_{1:t})$. For individual-level scores, inference requires access to the latent variables, and therefore targets the augmented posterior $\pi(\Theta \mid Y_{1:t})$. In this case, particles must include both parameters and latent variables.

\subsubsection{IBIS for benefit--risk analysis}

When the likelihood can be expressed as $f(Y \mid \theta)$, the IBIS algorithm provides a sequential approximation of $\pi(\theta \mid Y_{1:t})$. The method maintains a set of weighted particles $\{(\omega_m,\theta_m)\}_{m=1}^{N_\theta}$, updated recursively as new data arrive. At each time $t$, particle weights are updated using the incremental likelihood contribution,
\[
u_t(\theta_m) = f(y_t \mid \theta_m),
\]
followed by normalisation. Posterior expectations of functions $g(\theta)$ can then be approximated by weighted averages,
\[
\frac{\sum_m \omega_m g(\theta_m)}{\sum_m \omega_m}.
\]

When MCDA scores depend on latent variables, each particle $\theta_m$ can be extended by sampling latent quantities from $\pi(\mathbf{z}_{1:t}, \mathbf{u}_{1:t} \mid Y_{1:t}, \theta_m)$, yielding particles in the augmented space. As is standard in SMC, weight degeneracy is monitored using the effective sample size (ESS),
\begin{equation}
\mathrm{ESS}(\omega) = \frac{\left(\sum_{m=1}^{N_\theta} \omega_m \right)^2}{\sum_{m=1}^{N_\theta} \omega_m^2},
\end{equation}
and resampling is triggered when ESS falls below a threshold. A rejuvenation step, implemented via MCMC targeting $\pi(\Theta \mid Y_{1:t})$ using the MCMC scheme described in the Supplementary Section S1, is then used to maintain particle diversity while preserving the target distribution. This implementation follows \citeA{Chopin2002a}, adapted to the benefit--risk setting through the sequential evaluation of MCDA scores.

\subsubsection{Extension to latent variable models}

For the mixed-type models of Section \ref{sec:models}, latent variables cannot be integrated out analytically, and inference must be performed on the augmented space $\Theta = (\theta, \{\mathbf{z}_i,\mathbf{u}_i\}_{i=1}^N)$. In this setting, a direct application of IBIS may be inefficient, as sampling latent variables from their prior leads to highly variable importance weights. We therefore adopt an extension of IBIS in which latent variables are sampled from an approximation to their conditional posterior. Specifically, for each observation $y_t$ and particle $\theta_m$, we draw latent variables from a proposal approximating $\pi(\mathbf{z}_t,\mathbf{u}_t \mid y_t, \theta_m)$, centred at the conditional posterior mode and with covariance determined by the local Hessian; that is, a Laplace approximation (details in the Supplementary Section S3).

This leads to a proposal distribution $q(\mathbf{z}_t,\mathbf{u}_t \mid y_t,\theta_m)$ that is better aligned with the target posterior. The corresponding importance weights are adjusted to account for the proposal,
\[
u_t(\theta_m, \mathbf{z}_t^{(m)}, \mathbf{u}_t^{(m)}) =
\frac{f(y_t \mid \theta_m, \mathbf{z}_t^{(m)}, \mathbf{u}_t^{(m)}) \pi(\mathbf{z}_t^{(m)}, \mathbf{u}_t^{(m)} \mid \theta_m)}
{q(\mathbf{z}_t^{(m)}, \mathbf{u}_t^{(m)} \mid y_t, \theta_m)}.
\]
Particles are then propagated in the augmented space, yielding approximations to $\pi(\Theta \mid Y_{1:t})$ for all $t$. Unlike generic SMC$^2$ approaches, the proposed scheme avoids nested particle systems and is computationally attractive for the present models. The scheme is summarised in Algorithm \ref{alg:modibis}.
\begin{algorithm}
\caption{Sequential IBIS with latent variables for benefit--risk analysis}
\label{alg:modibis}
\textbf{Initialisation:} Sample $\{\theta_m\}_{m=1}^{N_\theta} \sim \pi(\theta)$ and set $\omega_m = 1$.

\textbf{For} $t = 1,\dots,N$:
\begin{algorithmic}[1]
\State For each particle $m$:
\begin{itemize}
\item Sample latent variables $(\mathbf{z}_t^{(m)}, \mathbf{u}_t^{(m)})$ from a proposal distribution 
\[
q(\mathbf{z}_t, \mathbf{u}_t \mid y_t, \theta_m),
\]
constructed via a Laplace approximation to $\pi(\mathbf{z}_t, \mathbf{u}_t \mid y_t, \theta_m)$.
\item Compute incremental weight:
\[
u_t^{(m)} =
\frac{f(y_t \mid \theta_m, \mathbf{z}_t^{(m)}, \mathbf{u}_t^{(m)}) \,
\pi(\mathbf{z}_t^{(m)}, \mathbf{u}_t^{(m)} \mid \theta_m)}
{q(\mathbf{z}_t^{(m)}, \mathbf{u}_t^{(m)} \mid y_t, \theta_m)}.
\]
\end{itemize}

\State Update weights: $\omega_m \leftarrow \omega_m \, u_t^{(m)}$.

\State Compute $\mathrm{ESS}(\omega)$.

\If{$\mathrm{ESS}(\omega) < \gamma N_{\theta}$}
\State Resample particles $\{(\theta_m, \mathbf{z}_{1:t}^{(m)}, \mathbf{u}_{1:t}^{(m)})\}$.
\State Apply MCMC rejuvenation targeting $\pi(\Theta \mid Y_{1:t})$, using a short HMC-based rejuvenation move described in the Supplementary Section S1.
\EndIf

\State Use weighted particles to approximate expectations under $\pi(\Theta \mid Y_{1:t})$, including MCDA scores $M_r(\Theta)$.
\end{algorithmic}
\end{algorithm}

Here $\gamma$ denotes the effective sample size threshold used to trigger resampling; throughout, we set $\gamma = 0.5$. This scheme provides a sequential approximation of $\pi(\Theta \mid Y_{1:t})$ that is well suited to mixed-type data, while enabling direct propagation of uncertainty to the MCDA scores.

\subsection{Implications for dynamic benefit--risk assessment}

The sequential framework provides a dynamic view of benefit--risk evaluation by tracking the evolution of posterior distributions of MCDA scores as data accumulate. This allows assessment not only of point estimates, but also of the uncertainty surrounding treatment comparisons over time. In particular, the sequence $\{\pi(M_r(\Theta) \mid Y_{1:t})\}_{t=1}^N$ enables monitoring of treatment rankings and posterior probabilities of superiority as evidence accrues. This is especially relevant in clinical settings where decisions may need to be updated as new data become available. In Section \ref{sec:casestudy}, we illustrate how the proposed methodology can be used to track benefit--risk profiles sequentially as data accumulate, by examining how conclusions would have evolved had the analysis been conducted over time.

\section{Rosiglitazone Case Study}\label{sec:casestudy}

We illustrate the proposed framework using data from a clinical trial on Rosiglitazone for type 2 diabetes. We first describe the data and MCDA setup. We then apply the model assessment framework of Section \ref{section:modelfit} to select an appropriate latent factor model. Based on the selected model, we compute MCDA scores and treatment comparisons. Finally, we demonstrate the sequential methodology by reconstructing a hypothetical data arrival process and examining how conclusions would have evolved over time.

\subsection{Data and MCDA Setup}\label{subsec:mcda}

We analyse data from a clinical trial in which three treatments were administered over a 12-week period: Metformin (MET), Rosiglitazone (RSG), and their combination (AVM, marketed as `Avandia'). The sample sizes were 146, 153, and 150 subjects respectively. For each subject, the data consist of two continuous efficacy outcomes (changes in haemoglobin and glucose levels from baseline) and four binary adverse event indicators (diarrhoea, nausea, vomiting, and dyspepsia). MCDA parameters, including outcome ranges and weights, were specified in consultation with domain experts. These are summarised in Table \ref{table:mcda_params}.

\begin{table}[!htbp]
\centering
\begin{tabular}{cccc}
\toprule
Name & Type & Outcome Range & MCDA Weight \\
\midrule
haemoglobin         & continuous    & [-6, 3]          & 0.592 \\
glucose             & continuous    & [-15, 7.5]       & 0.118 \\
prob(diarrhoea)     & binary        & [0.10, 0.35]     & 0.089 \\
prob(nausea)        & binary        & [0.10, 0.25]     & 0.178 \\
prob(vomiting)      & binary        & [0.10, 0.20]     & 0.018 \\
prob(dyspepsia)     & binary        & [0.10, 0.25]     & 0.005 \\
\bottomrule
\end{tabular}
\caption{MCDA outcome definitions and weights.}
\label{table:mcda_params}
\end{table}

\subsection{Model Choice}\label{subsec:model}

We consider a range of candidate models and apply the assessment framework of Section \ref{section:modelfit}. Combined log-scores were computed using sequential out-of-sample prediction, with each observation scored using the predictive distribution based on all previously observed data, and the resulting log densities were aggregated across observations. The candidate models include the saturated (SAT) and independence (IND) specifications, as well as confirmatory factor models with exact zero (EZ) or approximate zero (AZ) loading structures, as defined in Section \ref{sec:models}. The suffix “c” denotes models with correlated latent factors, while “a” and “b” distinguish different approximate zero specifications, corresponding to models without and with cross-loadings, respectively. The suffix “-p” denotes versions in which covariance parameters are pooled across treatment groups, while models without this suffix allow group-specific covariance structures.

The results, summarised in Table \ref{table:cvscores-combined}, indicate that pooling covariance parameters improves predictive performance, suggesting similar dependence structures across treatment groups. The saturated models underperform relative to more parsimonious specifications, suggesting overfitting. Among factor models, independence between factors is preferred, as EZ-p outperforms EZc-p. Overall, the most parsimonious model, EZ-p, achieves the best predictive performance.

\begin{table}[!htbp]
\centering
\begin{tabular}{cc}
\toprule
Model & Combined Log-Score \\
\midrule
SAT     & 2,344.11 \\
SAT-p   & 2,339.14 \\
EZ      & 2,335.68 \\
EZ-p    & 2,322.25 \\
EZc-p   & 2,324.49  \\
AZa-p   & 2,324.51 \\
AZb-p   & 2,324.52 \\
IND     & 2,419.31  \\
IND-p   & 2,418.11 \\
\bottomrule
\end{tabular}
\caption{Out-of-sample predictive performance (lower is better).}
\label{table:cvscores-combined}
\end{table}

We further assess whether the models achieve satisfactory fit using PPP values using PPP values (Table \ref{table:ppp}), with values close to 0.5 for both continuous and binary outcomes.

\begin{table}[!htbp]
\centering
\begin{tabular}{ccc}
\toprule
Model & Continuous-PPP & Binary-PPP \\
\midrule
EZ-p   & 0.49  & 0.36  \\
EZc-p  & 0.49  & 0.37  \\
AZa-p  & 0.42  & 0.35  \\
AZb-p  & 0.44  & 0.39  \\
\bottomrule
\end{tabular}
\caption{Posterior predictive p-values for selected models.}
\label{table:ppp}
\end{table}

Parameter estimates for the selected model are reported in the Supplementary Section S4.

\subsection{MCDA Scores}\label{subsec:scores}

\begin{figure}
\centering
\includegraphics[width=1.0\textwidth]{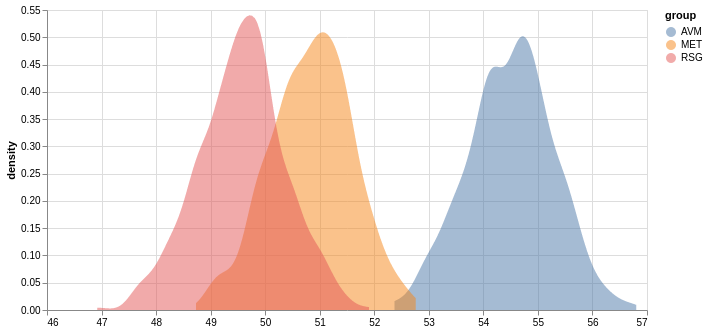}
\caption{Posterior distributions of MCDA scores for the three treatments. AVM dominates both MET and RSG.}
\label{fig:fig1}
\end{figure}

We compute posterior distributions of the MCDA scores using samples from the fitted model. For continuous outcomes, expectations are directly given by the intercept parameters, while for binary outcomes expectations are computed via Monte Carlo integration over latent variables. Applying this procedure to the selected model EZ-p yields posterior distributions of MCDA scores for the three treatments, shown in Figure \ref{fig:fig1}. AVM achieves higher posterior MCDA scores than both MET and RSG. The posterior probabilities of superiority are $P(M_{\text{AVM}} > M_{\text{MET}}) = 0.99$ and $P(M_{\text{AVM}} > M_{\text{RSG}}) = 0.99$.

\subsection{Sequential Analysis}\label{subsec:sequential}

To illustrate the sequential methodology, we reconstruct a hypothetical data arrival process by randomly ordering observations and interleaving treatment groups across treatment arms. Repeating the analysis under alternative random orderings produced qualitatively similar trajectories. This allows us to examine how inference would have evolved had the analysis been conducted sequentially. Under the selected model EZ-p, the posterior can be factorised into components corresponding to continuous and binary outcomes. The sequential algorithm is applied using the modified IBIS scheme (Algorithm \ref{alg:modibis}), with latent variables introduced only for the binary component. Details of the Laplace approximation used for latent variable proposals are provided in the Supplementary Section 3.

\begin{figure}
\centering
\includegraphics[width=1.0\textwidth]{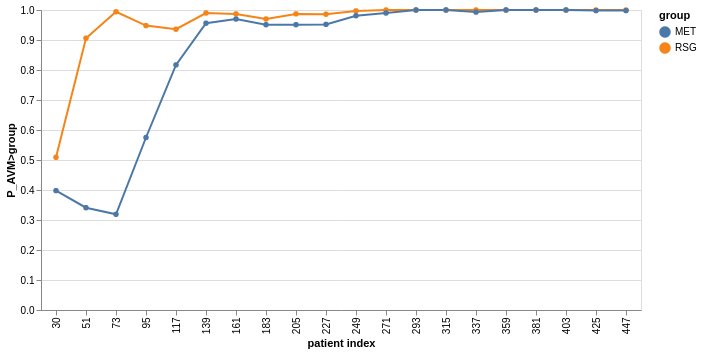}
\caption{Sequential evolution of posterior probabilities of treatment superiority.}
\label{fig:fig2}
\end{figure}

We analyse the data sequentially, updating posterior distributions after each observation. Figure \ref{fig:fig2} shows the evolution of the posterior probabilities $P(s_{\text{AVM}} > s_{\text{MET}})$ and $P(s_{\text{AVM}} > s_{\text{RSG}})$. These probabilities converge to 0.99 after approximately 200 and 300 observations respectively, indicating that strong evidence in favour of AVM emerges well before the full sample is observed. This illustrates how sequential analysis can provide insight into the rate at which evidence accumulates and suggests that, under a sequential design, strong conclusions might have been reached earlier. The relatively smooth evolution of these probabilities, without pronounced fluctuations, provides additional reassurance that the conclusions are not driven by transient features of the data sequence.

\begin{figure}
\centering
\includegraphics[width=1.0\textwidth]{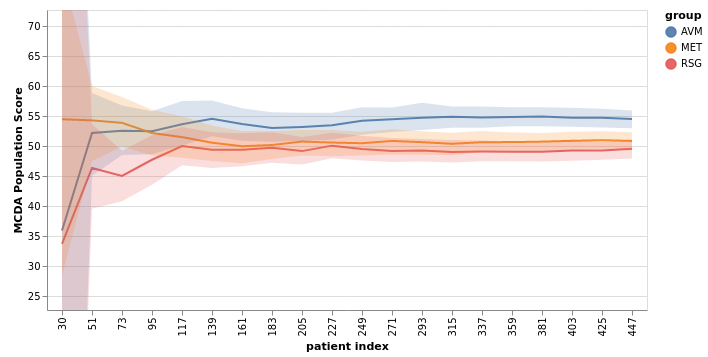}
\caption{Sequential evolution of posterior means and 95\% credible intervals for MCDA scores.}
\label{fig:fig4}
\end{figure}

Figure \ref{fig:fig4} presents the evolution of posterior means and 95\% credible intervals for MCDA scores. While AVM shows higher expected scores early in the trial, uncertainty remains substantial initially. Clear separation between treatments emerges after approximately 300 observations, at which point AVM is consistently superior.

\section{Discussion}

In this paper, we develop a Bayesian framework for benefit--risk assessment of clinical treatments that combines multi-criteria decision analysis with latent variable models for mixed-type outcomes. A key advantage of the MCDA framework is that it allows multiple efficacy and safety outcomes to be synthesised into a single treatment-specific score, while also enabling stakeholders---including clinicians, policymakers, and patients---to express preferences through the use of weights. The resulting MCDA scores, however, depend critically on the accurate estimation of the underlying statistical quantities implied by the underlying model.

To support this goal, we consider a range of models for mixed-type data together with a framework for model assessment and selection. This is important because richer models can capture more complex features of the data, including dependence across outcomes, but may also be more susceptible to overfitting. Conversely, overly restrictive models may fail to capture clinically relevant structure. Careful model assessment is therefore essential before posterior summaries are carried forward to the benefit--risk analysis. In the mixed-data setting considered here, goodness-of-fit and predictive performance cannot easily be summarised by a single joint criterion that simultaneously captures dependence across continuous and binary outcomes. In the absence of such a metric, and in line with \citeA{moustaki1996latent}, we assess performance separately by data type and require satisfactory behaviour for both components. Developing joint assessment criteria for mixed outcomes remains an important topic for future research.

We also propose a sequential framework for benefit--risk analysis that allows treatment comparisons and MCDA scores to be updated as data accumulate. The proposed framework is related to the broader literature on Bayesian adaptive clinical trial design, where posterior updating is used to support interim monitoring, treatment comparison, and resource allocation decisions; see, e.g., \cite{Berry2010}. 
Although our case study is retrospective, it illustrates how such a framework could be used to examine how conclusions would have evolved had the analysis been conducted sequentially. This may be particularly useful in settings where interim learning is important for monitoring treatment performance or adapting study design. The current implementation relies on the modelling assumptions adopted in the paper, including Gaussian latent variable structures and linear factor specifications. Extending the framework to more general non-Gaussian or non-linear settings would be a natural direction for future work.

An alternative approach to model comparison would be to use model evidence, which can be estimated naturally as a by-product of the sequential methodology. We instead focus on predictive performance, as our primary objective is not to identify a ``true'' model, but to select a model that is most useful for the downstream task of benefit--risk evaluation. From that perspective, predictive performance offers a more directly relevant criterion, as it rewards models that generalise well and yield reliable estimates of the treatment-specific quantities entering the MCDA scores.

\section*{Acknowledgments}
We thank GlaxoSmithKline Research \& Development Ltd for providing access to the clinical trial data via `www.clinicalstudydatarequest.com'. This publication is based on research using data from GlaxoSmithKline that has been made available to us through secured access. CSDR team or GSK team has not contributed to or approved, and is not in any way responsible for, the contents of this publication.

\bibliography{references}

\vspace{\fill}

\end{document}